# PHILOSOPHY OF SCIENCE VIEWED THROUGH THE LENSE OF "REFERENCED PUBLICATION YEARS SPECTROSCOPY" (RPYS)

K. Brad Wray** & Lutz Bornmann*

** Department of Philosophy

State University of New York, Oswego

212 Campus Center

Oswego, New York

13126, United States of America

brad.wray@oswego.edu

* Division for Science and Innovation Studies,

Administrative Headquarters of the Max Planck Society,

Hofgartenstr. 8,

80539 Munich, Germany.

E-mail: bornmann@gv.mpg.de

**Abstract**

We examine the sub-field of philosophy of science using a new method developed in information science, Referenced Publication Years Spectroscopy (RPYS). RPYS allows us to identify peak years in citations in a field, which promises to help scholars identify the key contributions to a field, and revolutionary discoveries in a field. We discovered that philosophy of science, a sub-field in the humanities, differs significantly from other fields examined with this method. Books play a more important role in philosophy of science than in the sciences. Further, Einstein's famous 1905 papers created a citation peak in the philosophy of science literature. But rather than being a contribution to the philosophy of science, their importance lies in the fact that they are revolutionary contributions to physics with important implications for philosophy of science.

**Keywords**

citations; philosophy of science; publications; citation peaks; Thomas Kuhn; Albert Einstein; Karl Popper; logical empiricism

**INTRODUCTION**

Citation counts are used as an indicator to measure the impact of publications in the scientific community. This common use of citation counts is based on the fact that research usually builds on previous investigations, discussions, and discoveries in a scientific community. "Original ideas seldom come entirely 'out of the blue'. They are typically novel combinations of existing ideas" (Ziman 2000, 212; see also Simonton 2004). According to Robert K. Merton (1965), scientists build their research on the shoulders of giants, the great scientists of the past. Earlier findings are recombined and developed, resulting in the accumulation of knowledge, which in turn contributes to scientific progress. New knowledge cannot be generated without this relationship with past research.

As a rule, the relationship to earlier publications is expressed in the form of references. One can expect that the content of the earlier publication, the cited source, and that of the later publication, the citing source, are related and that the former is of significance to the knowledge claims in the latter. The premise of a normative theory of citations is that the more frequently scientific publications are cited, the more important they are for the advancement of knowledge (Bornmann & Daniel 2008). From this perspective, citation data provides interesting insights into the historical context of science (Garfield 1963).

We believe that the investigation of the historical context of a scientific field would be enhanced by employing a new methodology. With the methodology we employ, the citation analysis starts by selecting all papers dealing with a specific research topic or field. Then all cited references are extracted from the papers of this field-specific publication set and we analyze which papers have been referenced (cited) most often. This method is known as cited reference analysis (see Bornmann & Marx 2013). The focus on the important historical publications in a specific research field is a special application of the cited reference analysis. An analysis with specific emphasis on the publication years of the references can be used to measure the significance of key publications in a field and to reveal the historical roots of a given research field. This application of the cited reference analysis was introduced by Marx, Bornmann, Barth, & Leydesdorff (2014) as "Referenced Publication Years Spectroscopy" (RPYS) and used by Barth, Marx, Bornmann, and Mutz (2014) in physics, by Leydesdorff, Bornmann, Marx, and Milojevic (2014) in information science and by Marx and Bornmann (2014) in biology.

We propose to examine the sub-field of philosophy of science with RPYS. Philosophy of science is an interesting case study because it is not a scientific field, strictly speaking. It is a sub-field of philosophy which has a long tradition of being identified as part of the humanities, grouped with the fields of history and literature. No doubt the various fields in the humanities have different citation practices from the various fields in the natural sciences. First, books continue to play a more important role in the humanities than in the natural sciences. Though Newton's *Principia* was a crucial publication in physics in the late 17$^{th}$ century, few physicists publish their research findings in books these days. Philosophers, on the other hand, continue to publish books as well as articles. Second, unlike the sciences the humanities treat texts as objects of study. For example, Shakespeare's *Tempest* has generated a vast scholarly literature, as the play is subjected to ongoing analyses. In contrast, in the sciences, texts, specifically articles, are merely treated as means to convey knowledge. They are not themselves the objects of scientific study. Indeed, earlier this fact was noted by Merton. As Merton explains, "discriminating citation-studies must ... distinguish between citations to research studies and to 'raw data' — i.e. historical documents, poems and other literature of the distant past which humanists critically re-examine" (Merton 1968, 29, Note 57).

We have limited our data set to four key journals in the sub-field of philosophy of science: *British Journal for the Philosophy of Science*, *Erkenntnis*, *Philosophy of Science*, and *Studies in History and Philosophy of Science*. These are recognized as four of the six key journals in the sub-field (see Wray 2010). We have not included the other two journals identified as core journals in philosophy of science by Wray (2010), *Journal of Philosophy* and *Synthese*, because they both have a broader scope than the philosophy of science, and would thus introduce a lot of noise into our data set created by citations to papers that have no bearing on philosophy of science. We are also limited in the specific years of publications we include in our data set by the source from which we are drawing our data, specifically the *Web of Science* (WoS, Thomson Reuters).

**METHODS**

RPYS reveals the historical papers most relevant to the evolution of a research field. However, their historical role can only be determined through a careful analysis by experts in the field under study.

We started by selecting all papers (n=8,757 records, all document types) published in the journals *Philosophy of Science*, *British Journal for the Philosophy of Science*, *Studies in History and Philosophy of Science*, and *Erkenntnis* (date of search: October 2013) (see Table 1). The search was undertaken in the WoS – a multi-disciplinary literature database – and was not restricted to a specific time period. Since papers published in *Erkenntnis* have been covered in WoS only since 2000, the early papers from this journal could not be included in this study. In the second step, we extracted all cited references (n=156,560) from this publication set for further analyses. We were particularly interested in the referenced publication years (RPYs) in which particular publications (and authors) have been frequently cited.

Quantitative analysis of the publication years of all the publications cited in the publications in one specific research field generally shows that RPYs lying further back in the past are not represented equally, but that some RPYs appear particularly frequently in the references (Marx, Bornmann, Barth, &Leydesdorff, 2014). These frequently occurring RPYs become more differentiated towards the past and show up as distinct peaks in the RPY distribution curves. If one analyzes the publications underlying these peaks, it is possible to see that during the 19th and the first half of the 20th century they are predominantly formed by single relatively highly cited publications. As a rule these few particularly frequently cited publications contain the historical roots to the research field in question and it is possible to determine how the relationship to earlier publications developed over time. Towards the present, the peaks are not the results of single publications. Due to the many publications cited in the more recent RPYs, the proportion of individual much-cited publications in the RPYs falls steadily.

The analysis of the RPYs is not a new bibliometric approach but has already been discussed by De Solla Price (1974). It was applied by Van Raan (2000) to measure the growth of science and to detect important breakthroughs in science without pre-defining any field. The RPYS is also based on the citation-assisted background (CAB) method. CAB has been proposed by Kostoff and Shlesinger (2005) and is a "systematic approach for identifying seminal references" (p. 199) in a specific field.

The program RPYS.exe

The program RPYS.exe can be used to generate a RPYS of any set downloaded from the WoS. The procedure for how to use the routine is described in detail at http://www.leydesdorff.net/software/rpys/. Based on WoS data as input (for example, papers published in the four journals used here), the program generates two output files: "rpys.dbf" and "median.dbf."

"Rpys.dbf" organizes the number of cited references per referenced publication year. This file can be used in Excel for drawing a spectrogram of the data. "Median.dbf" contains the deviation of the number of cited references in each year from the median for the number of cited references in the two previous, the current, and the two following years ($t-2$; $t-1$; $t$; $t+1$; $t+2$). This deviation from the five-year median provides a curve smoother than the one in terms of absolute numbers. Both curves can be visualized using median.dbf (e.g., in Excel).

**RESULTS**

The aggregate data for philosophy of science scholarship exhibits a number of peaks between 1900 and 1970, a span of 71 years.  As the deviations from the median in Figure 1 show, the highest peaks appear in 1905, 1949, 1950, 1957, 1962, 1965, and 1970.  In our analysis, we will focus narrowly on these seven peaks.  We believe the seven highest peaks are likely to bring to our attention some of the most significant contributions to the field of philosophy of science in the past 71 years.Let us consider each of the seven peaks in sequential order, from the earliest peak in 1905 to the most recent high peak in 1970.

First peak in 1905:

Let us start with the first peak in the philosophy of science scholarship since 1900, the 1905 peak. The author who is cited most frequently in 1905 is not a philosopher of science, but a scientist, namely Albert Einstein. In fact he is responsible for 24 % of the citations to work published in 1905, an astounding proportion of the citations to works published in 1905.If this were a typical scientific field, one would expect to see the most cited work is a foundational study in the field published at that time. But this is not the case in philosophy of science. Philosophy of science is part of philosophy which in turn is a part of the humanities. As noted above, in the humanities, unlike the natural sciences, publications are not only vehicles of new knowledge. They are also objects of study.

1905 is the year in which Einstein published his foundational work in relativity theory. This work is not itself a contribution to the philosophy of science. Hisarticles were not written to be contributions in philosophy of science. Rather, physicists are the intended audience of Einstein's highly cited 1905 publications.  The most cited of Einstein's 1905 publications in our data set is "ZurElektrodynamikbewegterKörper" ("On the Electrodynamics of Moving Bodies").  This is the foundational paper dealing with the theory of Special Relativity.  55 % of the citations to Einstein's 1905 publications are to this paper.

Einstein's articles, and especially the paper on the electrodynamics of moving bodies, are objects of great interest to philosophers of science, and especially philosophers of physics. The members of the Vienna Circle, the group that is largely identified as the forerunners of contemporary philosophy of science, were inspired to a great extent by Einstein's work (see Frank 1949, pages 19-21). It created new and fascinating problems for the philosophy of science. Einstein's discoveries led to an important revolutionary change of theory, the overthrow of what was regarded as the greatest scientific achievement to date, Newton's physical theory. There were other roughly contemporary developments that contributed to interest in the philosophy of science at the beginning of the 20th Century, like the development of non-Euclidean geometries in the 19th Century (see Frank 1949 pages 12-13 and 21-23). But Einstein's work became a focal point for philosophers of science.  So, in addition to creating a set of problems that would occupy the attention of philosophers of physics, the revolutionary change of theory in physics led philosophers of science in general to think differently about scientific progress, and the nature of theories.  In the second edition of *The Routledge Companion to Philosophy of Science,* a highly respected and widely used reference source for philosophers of science, the Index entry on Einstein runs five lines, and includes sub-headings of "Bayesianism," "conventionalism," quantum mechanics," "space and time," "symmetry," and "theory change" (see Curd and Psillos 2014, 687).

Thus, citations to Einstein represent a vestige of the origins of philosophy of science in the humanities. Einstein's work is not a philosophical contribution, a contribution to the specialty literature itself. Rather, his articles are objects of study, but such crucial objects of study that they created a peak in the philosophy of science literature.

Incidentally, there was also important work published in 1905 by scholars who are now part of the canon in the philosophy of science, specifically Henri Poincaré and Pierre Duhem.  Poincaré was an accomplished mathematician and Duhem was an accomplished scientist, but they also made insightful contributions to the philosophy of science. Not surprisingly, their 1905 publications make no reference to Einstein and his contributions. Poincaré was the second most cited author publishing in 1905 in the philosophy of science literature. Citations to his work constitute 9.8 % of the citations to work published in 1905. Poincaré's work has been discussed at length in the realism/anti-realism debate, a perennial issue in the philosophy of science. He has been cast as the founder of structural realism. He did not use this term to describe his view. Rather, John Worrall (1989) is responsible for attaching the label to Poincaré.  Many contemporary papers on structural realism will cite Poincaré almost as a courtesy, a discussion of the specifics of his work is often not present.

Second peak in 1949:

The author whose publications from 1949 are most cited in our data set is Hans Reichenbach. Citations to Reichenbach's 1949 publications constitute 7.3 % of the citations to work published in 1949 in our data set. Reichenbach was a member of the Berlin School of logical empiricists, the Berlin equivalent of the Vienna Circle.  Indeed, he was one of the founding members of the group (see Rescher 2006).  Because of his Jewish ancestry Reichenbach was compelled to leave Germany when Hitler came to power.  By the 1940s he was in the United States of America, and was a leader in transplanting logical empiricism to America (see Reisch 2005).

Reichenbach's most commonly cited publication of 1949 was his book *The Theory of Probability*. This book was originally published in German in 1935 as *Wahrscheinlichkeitslehre : eine Untersuchung über die logischen und mathematischen Grundlagen der Wahrscheinlichkeitsrechnung*.  The 1949 edition includes significant revisions to the 1935 edition.  Reichenbach's place in this peak indicates the importance of both logical empiricism and probability theory in the history of philosophy of science. Furthermore, we see the importance of books in philosophy of science, thus underscoring the roots of philosophy of science in the humanities.  In a typical scientific field, the most cited sources are articles.

Third peak in 1950:

This peak is due to the contributions of the logical empiricists, but in this case the most cited author is Rudolf Carnap. Carnap is responsible for an impressive 14.4 % of the citations to work published in 1950 included in our data set. Carnap was part of the Vienna Circle of the 1920s and 1930s. He also fled Europe when the Nazis came to power. And he too contributed to the spread of logical empiricism in the United States. Carnap's most cited publication from 1950 is *The Logical Foundations of Probability*. Again, it is a book that is the most cited publication. It is worth mentioning that Carnap and Reichenbach founded the journal *Erkenntnis*, one of the key journals in philosophy of science.

<u>Fourth peak in 1957</u>:

The most highly cited author of works published in 1957 is Karl Popper. Popper's 1957 publications are responsible for about 6.2 % of the citations to works published in 1957 in our data set. Interestingly, there are a number of publications by Popper that are responsible for his high rate of citation. It is not a *single* article or book. But the most cited 1957 publication is *Poverty of Historicism* (a book).

<u>Fifth peak in 1962</u>:

This is the year that the first edition of Thomas Kuhn's *Structure of Scientific Revolutions* (in short "*Structure*") was published. Kuhn accounts for 6.6 % of the citations to publications from 1962 in our data set. The bulk of them are to the first edition of *Structure*, rather than other publications by Kuhn in 1962. Again it is a book rather than an article that is highly cited. As we will see, had there not been a second edition of *Structure* published in 1970, 1962 would be an even higher peak than it is. At one point, the second edition became the definitive edition to cite.

<u>Sixth peak in 1965</u>:

The most cited source in our data set from 1965 is Carl Hempel's *Aspects of Scientific Explanation* (in short "*Aspects*"). Hempel's 1965 publications are responsible for 15.2 % of the citations to publications from 1965 in our data set. Again it is a book that is the most cited publication. But it is worth mentioning that Hempel's *Aspects* is more aptly described as an anthology of (mostly) previously published papers. Hempel was part of the Berlin School (see Rescher 2006), and thus indicates the continuing importance of logical empiricism, even after the publication of Kuhn's *Structure*.

Seventh peak in 1970:

This is the highest peak in our data set. It is due, to a large extent, to the publication of the second edition of Kuhn's *Structure*. For over 15 years, until the publication of the 3rd edition in 1996, this was the definitive edition of *Structure* to cite. It included a new Postscript not included in the original 1962 publication of *Structure*, but of significant interest to philosophers of science. Kuhn is responsible for 10.5 % of the citations to works published in 1970.

Incidentally, the citation peak of 1970 is due in part to the fact that there are also many citations to work addressing Kuhn, including the various essays in *Criticism and the Growth of Knowledge*, a volume edited by Lakatos and Musgrave.

It is worth noting that because we restrict our analysis to data in the four core philosophy of science journals, we are not counting Kuhn's influence in other fields, which is enormous. In fact, *Structure* was the most cited source in the *Arts and Humanities Citation Index* between 1976 and 1983 (see Garfield 1987, 103). It was cited more frequently than Wittgenstein's *Philosophical Investigations*, Chomsky's *Aspects of the Theory of Syntax*, Foucault's *Order of Things*, and Heidegger's *Being and Time*, all highly cited and influential books (see Garfield 1987, 103-104).

The 1970 peak represents a genuine growth or turning point in philosophy of science. It would be misleading to say that there was a revolutionary change such that most philosophers of science adopted a Kuhnian perspective on science. It would be more accurate to claim that *Structure* became a central part of the disciplinary canon. It became a source of problems and questions, and thus gave a new focus for the sub-field. Gary Gutting, though, suggests that since the 1980s, Kuhn's influence in philosophy of science has been marginal (see Gutting 2009, 151). Gutting is probably overstating things. At any rate, a careful study of citations may shed light on this issue.

Discussion of the seven highest peaks:

We have examined the seven most important peaks in our data set and interesting anomalies arise in philosophy of science. In two of these peak years, the most cited author, Thomas Kuhn, is cited most frequently for a book rather than an article. Indeed, in all but one of the seven peaks it is a book that is the most cited source. Philosophy of science thus appears to be a book-driven discipline. In the first peak in the data set we analyzed, we encounter another anomaly. The most frequently cited author, Albert Einstein, is not even a philosopher, and his contributions are contributions to physics, not philosophy of science. His most cited articles from 1905 were very stimulating articles for philosophers thinking about science.

So the peaks in philosophy of science do not only appear to be a consequence of the publication of key *articles* in the philosophy of science. The seven highest peaks between 1900 and 1970 are tied to books and scientific articles that report scientific findings that are philosophically interesting. In this respect, philosophy of science is quite different from other fields, specifically fields in the natural sciences.

**DISCUSSION**

The RPYS method is based on an analysis of the frequency with which references are cited in the publications in a specific research field by publication year. The origins show up in the form of more or less pronounced peaks mostly caused by individual historical publications which are cited particularly frequently. The RPYS can only indicate the possible origins. Intellectual influences are not always manifest in cited references. A second step is required in which experts verify which publications genuinely played a significant part in a research field. When those publications which resulted in a peak are identified, each of them should be reviewed for their significance in the particular research field and what contribution they made. RPYS is a very simple method which can be applied in different disciplines.

We have employed RPYS in a novel way in this paper. Instead of using it to identify the most cited publications, we have concentrated on identifying the most cited authors. Further, this is, as far as we know, the first application of RPYS to a sub-field in the humanities.

Our findings alert us to differences in the ways various academic research cultures behave, as manifested in their publication and citation practices. Unlike the natural sciences, in philosophy of science books still play a crucial role in scholarship. Kuhn's *Structure of Scientific Revolutions* was partially responsible for not one, but two of the highest citation peaks in the scholarly literature in philosophy of science (1962 and 1970). Books were also responsible for the peaks in 1949, 1950, 1957, and 1965.

An additional anomaly in philosophy of science was identified with the first peak we examined. This peak was due to the many citations to articles published by Albert Einstein in 1905. These articles are not strictly speaking contributions to philosophy of science. But as important contributions to physics, they have become the objects of great interest to philosophers of science. Einstein ushered in revolutionary changes in physics, and such changes are of great interest to philosophers of science.

Our preliminary study is intended, in part, to showcase the value of this relatively new method, RPYS. Employing this method, we have also uncovered a number of interesting aspects of the epistemic culture of philosophy of science.

Alternative methods to the RPYS for analysing historical papers are the methods of co-citation analysis, bibliographic coupling, and direct citations in consideration of research fronts (Boyack & Klavans, 2010) and also a method called "algorithmic historiography" (Garfield, 2001; Garfield et al., 2003; Leydesdorff, 2010). The HistCite™ software developed by Alexander Pudovkin and Eugene Garfield (<http://garfield.library.upenn.edu/algorithmichistoriographyhistcite.html>) enables the establishment of the citation graph (sometimes called historiogram or historiograph) which visualizes the citation network among publication sets, including historical papers. The RPYS is simpler than the alternatives and does not reveal the citation network of the historical papers. The method only reveals quantitatively which historical papers are of particular interest for the specific research field or topic.

Table 1.Number of papers and number of cited references in four journals relevant for the philosophy of science.

| Journal | Number of papers | Number of cited references |
|---|---|---|
| *British Journal for the Philosophy of Science* (since 1956) | 2,814 | 35,985 |
| *Erkenntnis* (since 2000) | 714 | 15,973 |
| *Philosophy of Science* (since 1956) | 3,833 | 54,683 |
| *Studies in History and Philosophy of Science* (since 1974) | 1,396 | 49,919 |
| Total | 8,757 | 156,560 |

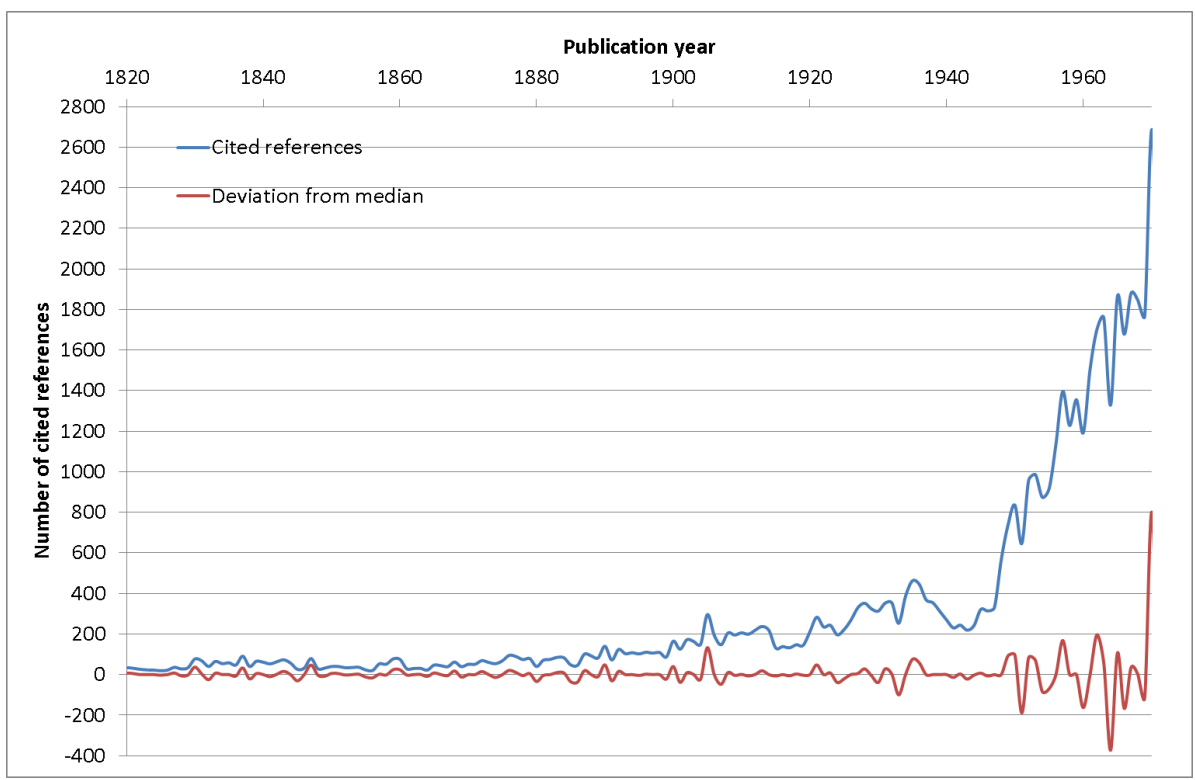

Figure 1. Referenced Publication Years Spectroscopy (RPYS) of four journals relevant for the philosophy of science: *British Journal for the Philosophy of Science*, *Erkenntnis*, *Philosophy of Science*, and *Studies in History and Philosophy of Science*.